\documentclass{pasj00}
\begin{document}
\SetRunningHead{J.\ Fukue}
{Relativistic Variable Eddington Factor}
%in Relativistic Radiative Flow}
\Received{yyyy/mm/dd}
\Accepted{yyyy/mm/dd}
%tex          2008 1010
%referee      2008 1202
%editing      

\title{Relativistic Variable Eddington Factor in a Relativistic Plane-Parallel Flow}
%in Relativistic Radiative Flow}

%%% begin:list of authors
\author{Jun \textsc{Fukue}} %   \thanks{}}
\affil{Astronomical Institute, Osaka Kyoiku University, 
Asahigaoka, Kashiwara, Osaka 582-8582}
\email{fukue@cc.osaka-kyoiku.ac.jp}

%\author{B-Firstname \textsc{B-Familyname}}
%\affil{B-Address of Institute}\email{bbbbb@xxx.xxx.xx.xx}
%\and
%\author{C-Firstname {\sc C-Familyname}}
%\affil{C-Address of Institute}\email{ccccc@xxx.xxx.xx.xx}
%%% end:list of authors

%% `\KeyWords{}' always has to be placed before `\maketitle'.
\KeyWords{
accretion, accretion disks ---
astrophysical jets ---
%black holes ---
%galaxies: active ---
gamma-ray bursts ---
%X-rays: individual (SS~433, GRS~1915$+$105, GRO~J1655$-$40) ---
radiative transfer ---
relativity
%X-rays: stars
} %Do NOT move this preamble from here!

\maketitle

%\newpage

\begin{abstract}
We examine the behavior of the variable Eddington factor 
for a relativistically moving radiative flow in the vertical direction.
We adopt the ``one-tau photo-oval'' approximation in the comoving frame.
Namely, the comoving observer sees radiation coming from
a closed surface where the optical depth measured from the observer is unity;
such a surface is called a {\it one-tau photo-oval}.
In general, the radiative intensity emitted by the
photo-oval is non-uniform and anisotropic.
Furthermore, the photo-oval surface has a relative velocity 
with respect to the comoving observer,
and therefore, the observed intensity suffers from the Doppler effect
and aberration.
In addition, the background intensity usually depends on the optical depth.
All of these introduce the {\it anisotropy}
to the radiation field observed by the comoving observer.
As a result,
the relativistic Eddington factor $f$ generally depends
on the optical depth $\tau$,
the four velocity $u$,
and the velocity gradient $du/d\tau$.
In the case of a plane-parallel vertical flow,
we found that the relativistic variable Eddington factor $f$ 
generally decreases as the velocity gradient increases,
but it increases as the velocity increases for some case.
When the comoving radiation field is uniform, 
it is well approximated by
$3f \sim 1/[ 1+
({16}/{15})(-{du}/{\gamma d\tau})
+(-{du}/{\gamma d\tau})^{1.6-2}
               ]$.
When the radiation field in the inertial frame is uniform,
on the other hand,
it is expressed as
$f = (1+3\beta^2)/(3+\beta^2$).
These relativistic variable Eddington factors can be used
in various relativistic radiatively-driven flows,
such as black-hole accretion flows,
relativistic astrophysical jets and outflows, and
relativistic explosions like gamma-ray bursts.
\end{abstract}

\section{Relativistic Variable Eddington Factor}

In the moment formalism of (relativistic) radiation hydrodynamics,
in order to close the moment equations truncated at a finite order,
we need a {\it closure relation},
such as the Eddington approximation or more complex expressions.
In non-relativistic static atmospheres
(Chandrasekhar 1960; Mihalas 1970; Rybicki \& Lightman 1979;
Mihalas \& Mihalas 1984; Shu 1991; Peraiah 2002; Castor 2004),
for example,
many researchers have used the Eddington approximation
and variable Eddington factors as closure relations
(Milne 1921; Eddington 1926; Kosirev 1934; Chandrasekhar 1934;
Hummer \& Rybicki 1971; Wilson et al. 1972; Tamazawa et al. 1975;
Unno \& Kondo 1976; Masaki \& Unno 1978).
As is well known, the standard Eddington approximation 
with the Eddington factor of $1/3$ is valid
when the radiation field is almost {\it isotropic}.
In the spherically symmetric case,
where the radiation field becomes anisotropic towards
the outer optically thin region,
they often used the variable Eddington factor,
which depends on the optical depth (e.g., Tamazawa et al. 1975).

In relativistically moving flows
(Mihalas et al. 1975; Mihalas et al. 1976a, b;
Thorne 1981; Thorne et al. 1981; Flammang 1982, 1984;
Mihalas \& Mihalas 1984; Nobili et al. 1991, 1993; 
Kato et al. 1998, 2008; Castor 2004; Mihalas \& Auer 2001;
Park 2001, 2006; Takahashi 2007), on the other hand,
the standard Eddington approximation is adopted in the comoving frame,
and then transformed to the inertial frame, if necessary
(Castor 1972; Hsieh \& Spiegel 1976; Fukue et al. 1985;
Sen \& Wilson 1993; Baschek et al. 1995, 1997; Kato et al. 1998, 2008).
Even for a comoving observer, however,
the comoving radiation field may become {\it anisotropic}
due to various reasons as shown below.
Instead of the standard Eddington approximation in the comoving frame,
for such a relativistic case
several types of a {\it variable Eddington factor},
which depends on the velocity and its gradient
as well as the optical depth, were proposed
(Fukue 2006; Fukue \& Akizuki 2006, 2007; Akizuki \& Fukue 2008;
Fukue 2008a, b; Koizumi \& Umemura 2008).
However,
the moment formalism of relativistic radiation hydrodynamics
is yet {\it incomplete},
in a sense that
we do not have the sufficiently adequate closure relation.

There are mainly three reasons,
which cause the anisotropy of the comoving radiation field
in the relativistically moving flow.
(1) The optical depth effect.
In the outer region where the flow becomes optically thin,
similar to the non-relativistic case,
the comoving radiation field would be anisotropic.
%In addition, we should carefully treat the optical depth
%for the relativistically moving media,
%since the length suffers from the Lorentz-Fitzgerald contraction.
(2) The velocity gradient effect.
When the radiative flow is accelerated up to the relativistic regime
and there is a strong velocity gradient in the direction of the flow,
the velocity fields as well as the density distribution
are no longer uniform even in the comoving frame.
Then, the comoving radiation field also becomes non-uniform.
Furthermore, due to the relative speed between the observer
and the radiation field,
the observed radiation field suffers from the Doppler effect.
As a result,
the comoving radiation field becomes anisotropic.
(3) The effect of the relativistic speed itself.
Although we do not know the radiative intensity in the inertial frame,
in the highly relativistic regime
it is generally affected by the strong aberration effect
when it is converted to the comoving frame.
As a result, except for some special cases,
the comoving radiative intensity generally has strong anisotropy.

Hence, the relativistic variable Eddington factor (RVEF) generally
depends on the optical depth, the flow velocity,
and the velocity gradient.
In Fukue (2006) and Akizuki and Fukue (2008),
they proposed such variable Eddington factors
 from physical view points
in the plane-parallel and spherical cases, respectively.
In Fukue (2008a, b), on the other hand,
he derived semi-analytically variable Eddington factors
for the plane-parallel flows in the vertical direction,
and found that the Eddington factor decreases
as the velocity-gradient increases.
In these studies (Fukue 2008a, b), however,
the treatments were rather restrictive;
e.g., the comoving radiation field is assumed to be uniform.
Thus, in the present paper 
we extend the previous work under the similar procedure
to examine the above three reasons more extensively.

In the next section
we describe the shape of the surface (one-tau photo-oval),
where the optical depth measured by the comoving observer is unity,
and derive the expression of the one-tau photo-oval.
In section 3
we numerically calculate
the comoving radiation field within the photo-oval,
and the relativistic variable Eddington factor
for several cases.
%In section 4,
%we briefly discuss the present result and compare with the results
%of previous work.
The final section is devoted to concluding remarks.

\section{One-Tau Photo-Oval and Photo-Vessel}

Let us suppose a relativistic radiative flow,
which is accelerated in the vertical ($z$) direction,
and a comoving observer,
who moves upward with the flow (figure 1).

\begin{figure}
  \begin{center}
  \FigureFile(70mm,70mm){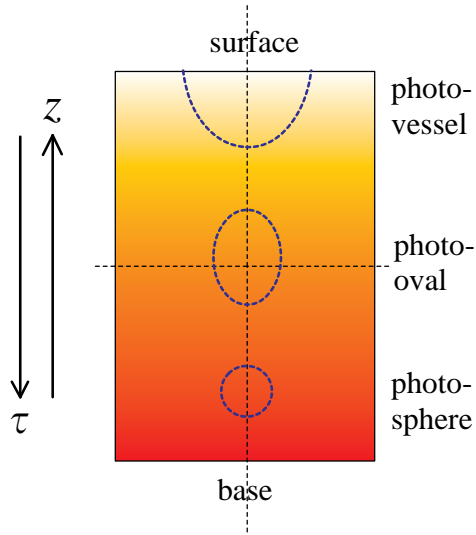}
  \end{center}
\caption{
Schematic picture of a relativistic radiative flow
in the vertical direction.
The flow is accelerated in the vertical ($z$) direction,
and has a velocity gradient.
The dashed curves are
one-tau photo-ovals observed by a comoving observer.
}
\end{figure}

Sufficiently deep inside the flow,
where the optical depth is very large,
the mean free path of photons is very short.
In such a case,
within the range of the mean free path,
for the comoving observer
the flow is seen to be almost uniform;
the velocity gradient and resultant density gradient are negligible.
Hence,
the mean free path is the same in all directions,
and the shape of the surface
where the optical depth measured from the comoving observer is unity,
is almost a sphere;
we call it a {\it one-tau photo-sphere} (a dashed circle in figure 1).
As a result,
the comoving radiation field is nearly isotropic,
and a usual Eddington approximation in the comoving frame is valid.

On the other hand,
if the velocity gradient becomes very large,
the behavior of the radiation field may be affected
by changes in the hydrodynamics of the flow
even on a lengthscale comparable to that 
of the photon mean free path.
Hence,
the mean free path becomes longer in the downstream direction
than in the upstream and other directions, and
the shape of the one-tau range elongates in the downstream direction;
we call it a {\it one-tau photo-oval} (a dashed oval in figure 1).
As a result,
the comoving radiation field becomes {\it anisotropic}
due to various reasons as described in the introduction,
and we should modify the usual Eddington approximation.

In addition,
when the optical depth is sufficiently small
and/or the velocity gradient is suffiently large,
the mean free path of photons in the downstream direction
become less than unity.
Hence,
the shape of the one-tau range is open in the downstream direction;
we call it a {\it one-tau photo-vessel} (a dashed hemi-circle in figure 1).

In order to obtain an appropriate form
of the relativistic variable Eddington factor in these regimes,
we thus carefully treat and examine
the radiation field in the comoving frame.

In previous papers (Fukue 2008a, b), 
we have examined the one-tau photo-oval,
derived the variable Eddington factor semi-analytically,
and proved that the Eddington factor decreases with the velocity gradient.
In these studies, however, the treatment were rather restricted,
and in the present study,
we thus examine the behavior of the relativistic variable Eddington factor
for more general cases,
and seek the appropriate form of the Eddington factor $f(\tau, u, du/d\tau)$,
where $\tau$ is the optical depth and
$u$ ($=\gamma \beta$) is the four velocity,
$\beta$ and
$\gamma$ ($=1/\sqrt{1-\beta^2}$) being 
the normalized velocity and the Lorentz factor, respectively.

\begin{figure}
  \begin{center}
  \FigureFile(80mm,80mm){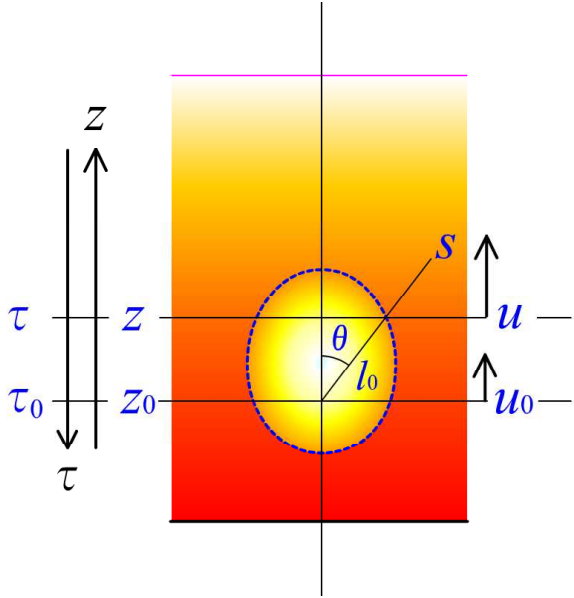}
  \end{center}
\caption{
One-tau photo-oval around a comoving observer
in the vertical ($z$) one-dimensional radiative flow.
The comoving observer is located at $\tau=\tau_0$,
where the flow four speed is $u=u_0$.
In the $s$-direction,
that forms an angle $\theta$ with the downstream direction,
the mean free path is set to be $l_0$.
}
\end{figure}

We first derive the shape of the one-tau photo-oval
using the optical depth $\tau$,
while in the previous papers we have used the vertical coordinate $z$.

We define the one-tau photo-oval as the surface
where the optical depth measured by the comoving observer is unity.
The situation is schematically illustrated in figure 2.
We assume that the comoving observer in the vertical flow is located 
at $z=z_0$ or $\tau=\tau_0$,
where the flow four speed is $u=u_0$.
In the $s$-direction,
that forms an angle $\theta$ with the downstream direction,
the mean free path of photons is $l_0$.
The relation among these quantities is
\begin{equation}
   z - z_0 = s \cos\theta.
\label{zz0}
\end{equation}

The continuity equation for the stationary, one-dimensional 
relativistic flow is
\begin{equation}
   \rho cu = J ~(={\rm const.}),
\label{rho}
\end{equation}
where $\rho$ is the proper gas density,
$u$ the four velocity,
and $J$ the mass-flow rate per unit area.
In addition, the optical depth $\tau$ is defined by
\begin{equation}
   d\tau \equiv -\kappa \rho dz = -\kappa \rho ds \cos\theta,
\label{tau}
\end{equation}
where $\kappa$ is the opacity,
which is assumed to be constant in the present analysis,
and we used equation (\ref{zz0}).

In this paper
we use a linear approximation for the flow field;
that is to say, around the position of the comoving observer
the flow four speed is expanded as
\begin{equation}
   u = u_0 + \left.\frac{du}{d\tau}\right|_0 ~(\tau-\tau_0),
\label{linear_u}
\end{equation}
where $du/d\tau|_0$ is assumed to be constant.

Using equation (\ref{tau}),
the optical depth $\tau_s$ along the $s$-direction
is expressed as
\begin{eqnarray}
   \tau_s &=& \int_0^{l_0} \kappa\rho ds
    = - \int_{\tau_0}^{\tau} \frac{d\tau}{\cos\theta}
    = \frac{\tau_0 - \tau}{\cos\theta}.
\label{tau_s}
\end{eqnarray}
Thus, the shape of the one-tau photo-oval, where $\tau_s=1$,
is finally determined by the condition:
\begin{equation}
\tau = \tau_0 - \cos\theta.
\label{oval}
\end{equation}
Obviously,
for $\tau_0>1$
the one-tau region is close to be a photo-oval,
whereas
it is open towards a downstream direction
to be a photo-vessel for $\tau_0<1$.

%%%%%%%%%%%%%%%%%%%%%%%%%%%%%%%%%%%%%%%%%%

\section{Comoving Radiation Field and the Relativistic Variable Eddington Factor}

We can now compute the radiation field 
received by the comoving observer,
and derive an expression for the Eddington factor
in the comoving frame (figure 3).

In a static and optically thick atmosphere,
the radiation field is isotropic and uniform.
In the present moving atmosphere, on the other hand,
there are three reasons for which the radiation field is 
not uniform and isotropic, as described in the introduction;
the optical depth effect, the velocity gradient effect,
and the effect of the relativistic speed itself.

First, the radiative intensity $I$ in the comoving frame
emitted from the one-tau photo-oval walls
%which is not known without solving the full transfer problem,
is generally a function of $\tau$ and $\mu$.
In general
the radiative intensity increases with the optical depth,
and therefore,
the intensity from the upstream direction is slightly
larger than that from the downstream direction
in the present one-dimensional vertical flow.
This effect of non-uniformity of intensity
generally acts as a force to accelerate a comoving observer.
This optical depth effect is prominent for small optical depth,
similar to the static atmosphere.

Second, although in the usual static atmosphere
the $\mu$-dependence is safely ignored for large optical depth,
this is not the present relativistic case.
Namely, in the relativistic flow
there appear the aberration and Doppler effects
between the inertial and comoving frames.
Hence, except for very special cases,
the comoving radiative intensity would have strong anisotropy.

\begin{figure}
  \begin{center}
  \FigureFile(80mm,80mm){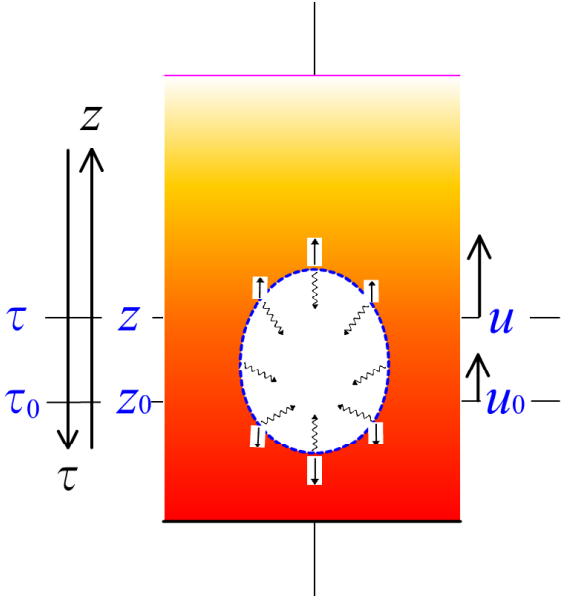}
  \end{center}
\caption{
Radiation field in the one-tau photo-oval around a comoving observer
in the vertical one-dimensional radiative flow.
The radiation field may become anisotropic,
since in general the emitted intensity is not uniform and isotropic,
it is redshifted due to the velocity difference,
and there is aberration between the comoving and inertial frames.
}
\end{figure}

Thirdly, if there is a velocity gradient in the flow,
the intensity observed by the comoving observer
is redshifted (Doppler shifted) due to the velocity difference between
the comoving observer and the one-tau photo-oval walls.
In an accelerating flow,
where the flow speed increases toward the downstream direction,
the relative velocity is generally positive (figure 3)
except for some special direction ($\theta=\pi/2$).
Hence, the Doppler shift of intensity also causes anisotropy
of the radiation field at the position of the comoving observer.

Neglecting the local emissivity,
the frequency-integrated intensity $I_{\rm co}$ observed in the comoving frame
at the position $\tau_0$ of the comoving observer is related to 
the frequency-integrated intensity $I(\tau, \mu)$ 
emitted at the one-tau photo-oval and
to the redshift $z$ by
\begin{equation}
   I_{\rm co} = \frac{I(\tau, \mu)}{(1+z)^4}.
\label{Ico}
\end{equation}
The relative speed $\Delta \beta$ between the comoving observer
and the one-tau photo-oval walls is given
by the relativistic summation law as
\begin{equation}
   \Delta \beta = \frac{\beta - \beta_0}{1-\beta\beta_0},
\label{Dbeta}
\end{equation}
where
\begin{eqnarray}
   \beta &=& \frac{u}{\sqrt{1+u^2}},
\\
   \beta_0 &=& \frac{u_0}{\sqrt{1+u_0^2}}.
\end{eqnarray}
It should be noted that
as long as the velocity gradient is positive
(accelerating flow)
the relative speed (\ref{Dbeta}) is always positive (or zero),
and the one-tau photo-oval is seen to expand (figure 3).
%
%In addition, using equations (\ref{semi_u0}) and (\ref{semi_l0final}),
%the four velocity at $z$ is expressed as
%\begin{equation}
%   u = u_0 e^{\left.-\frac{\displaystyle 1}{\displaystyle u}\frac{\displaystyle du}{\displaystyle d\tau}\right|_0 \displaystyle \cos\theta}.
%\end{equation}
%Hence, all the necessary quantities are expressed
%as functions of $u_0$, $du/d\tau|_0$, and $\cos\theta$.

Using the relative speed above,
the redshift $z$ is expressed as
\begin{equation}
   1+z = \frac{1+\Delta \beta \cos\theta}{\sqrt{1-(\Delta\beta)^2}}.
\end{equation}
%\begin{eqnarray}
%   I_{\rm co} &=& \frac{I_0}{(1+z)^4}.
%\label{RF_Ico}
%\end{eqnarray}

Now, all the quantities are calculated
for given $\tau_0$, $u_0$, $du/d\tau |_0$, and $I(\tau, \mu)$.
Once the observed intensity $I_{\rm co}$ is obtained by equation (\ref{Ico}),
the radiation energy density $E_{\rm co}$,
the radiative flux $F_{\rm co}$, and
the radiation pressure $P_{\rm co}$
measured by the comoving observer are calculated respectively as
\begin{eqnarray}
   cE_{\rm co} &\equiv& \int I_{\rm co} d\Omega_{\rm co},
\label{RF_Eco_linear}
\\
   F_{\rm co} &\equiv& \int I_{\rm co} \cos\theta d\Omega_{\rm co},
\label{RF_Fco_linear}
\\
   cP_{\rm co} &\equiv& \int I_{\rm co} \cos^2\theta d\Omega_{\rm co}.
\label{RF_Pco_linear}
\end{eqnarray}
Finally, the relativistic variable Eddinton factor 
in the comoving frame is given by
\begin{equation}
   f \equiv \frac{P_{\rm co}}{E_{\rm co}}.
\label{f_linear}
\end{equation}

In the following subsections,
we consider in turn several cases;
the velocity gradient effect,
the optical depth effect,
and the strong anisotropy of the radiative intensity.

\subsection{Uniform Case with Photo-Vessel}

We first consider the velocity gradient effect,
similar to the previous studies (Fukue 2008a, b),
and also the photo-vessel case,
which was not considered in the previous studies.
To clarify the velocity gradient effect,
we here assume that the comoving radiation field is uniform:
\begin{equation}
   I (\tau, \mu) = \bar{I} = {\rm const.}
\end{equation}

The numerical results of this uniform case
are shown in figure 4 for small $u_0$ and 
figure 5 for large $u_0$.
In figures 4a and 5a
the relativistic variable Eddington factors $f(u, -du/d\tau)$
multiplied by 3 are plotted in the parameter space,
whereas
they are plotted for several fixed $u_0$ with fitting curves
in figures 4b and 5b.

\begin{figure}
  \begin{center}
  \FigureFile(80mm,80mm){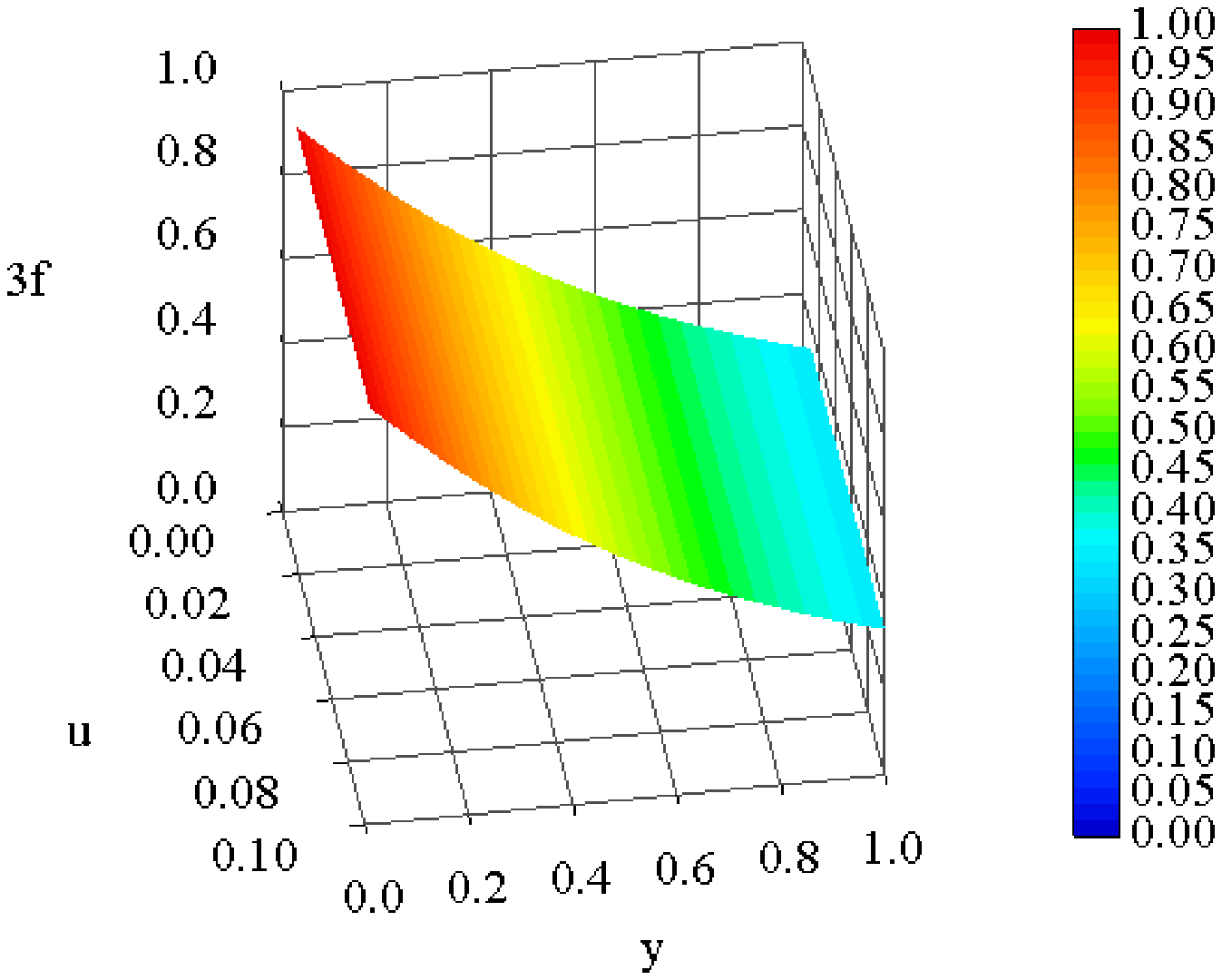}
  \FigureFile(80mm,80mm){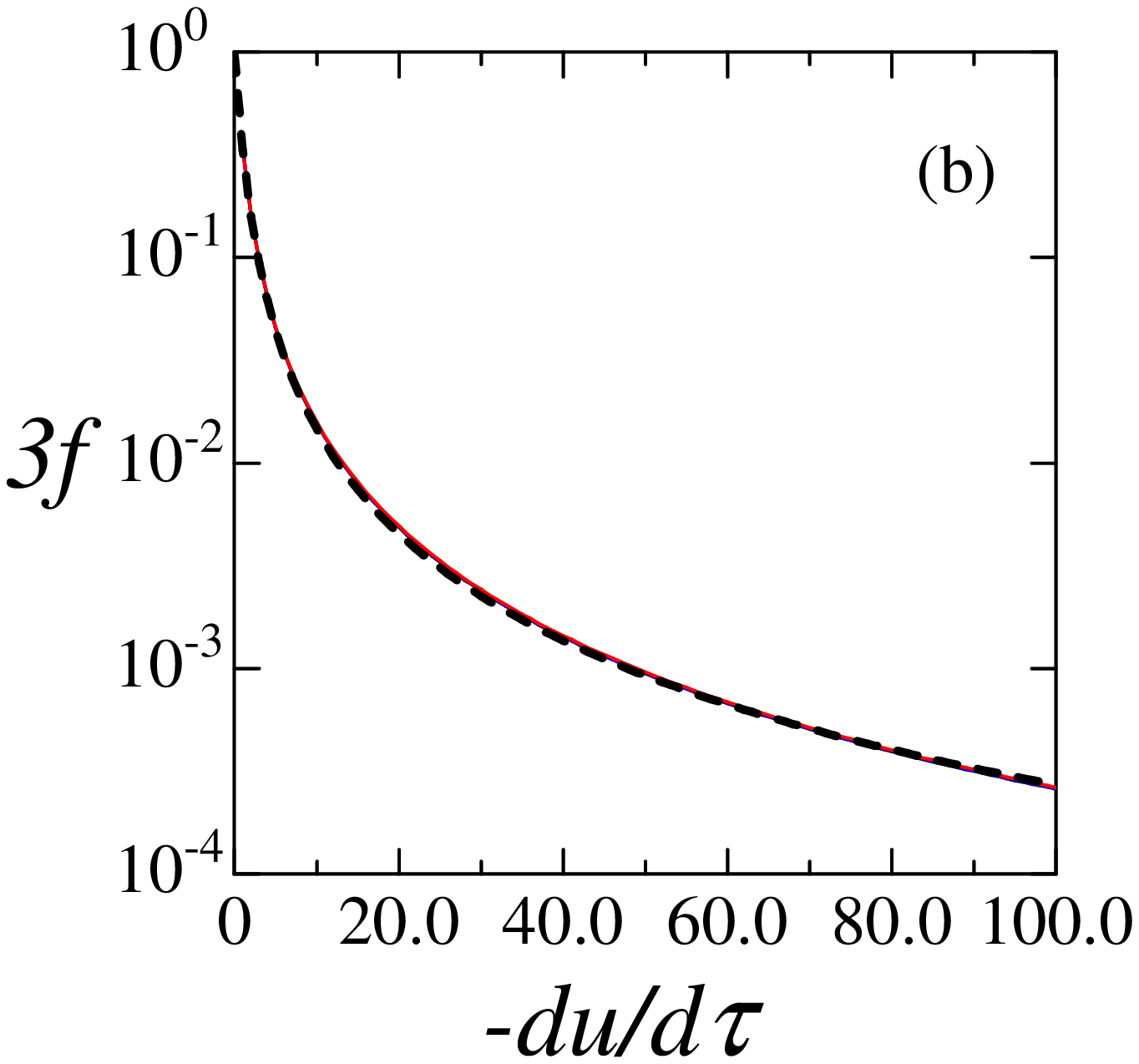}
  \end{center}
\caption{
Behavior of the relativistic variable Eddington factor $f(u, -du/d\tau)$
multiplied by 3 in the case of small $u$.
(a) $3f$ are plotted as a function of the four velocity $u_0$ ($x$ axis)
and the four-velocity gradient $-du/d\tau|_0$ ($y$ axis).
(b) $3f$ are plotted as a function of 
the four-velocity gradient $-du/d\tau|_0$
for four fixed $u_0$ ($=0.01, 0.05, 0.1$; solid curves),
and one fitting curve is also shown (dashed curve).
}
\end{figure}

The behavior of RVEF for small $u_0$ is shown in figure 4 (cf. Fukue 2008a).
In this case RVEF decreases,
as the velocity gradient becomes large, 
as analytically proved in Fukue (2008a).
In figure 4b for fixed $u_0$ ($=0.01, 0.05, 0.1$)
the values of $3f$ are plotted as a function of $-du/d\tau|_0$
by solid curves;
all curves overlap each other.
One fitting curve is also plotted
by a thick dashed one, and
the numerical results are well fitted by
\begin{equation}
   3f  \sim  \left( 1+\frac{16}{15}y + 0.9y^{1.8} \right)^{-1},
\end{equation}
where
\begin{equation}
y=-du/d\tau|_0,
\end{equation}
within an error of a few percent.
It should be noted that we chosed 
the coefficient, $16/15$, so  that
theis fitting equation reduces to
\begin{equation}
   3f = 1-\frac{16}{15}y
\end{equation}
in the linear regime as analytically proved in Fukue (2008a).

\begin{figure}
  \begin{center}
  \FigureFile(80mm,80mm){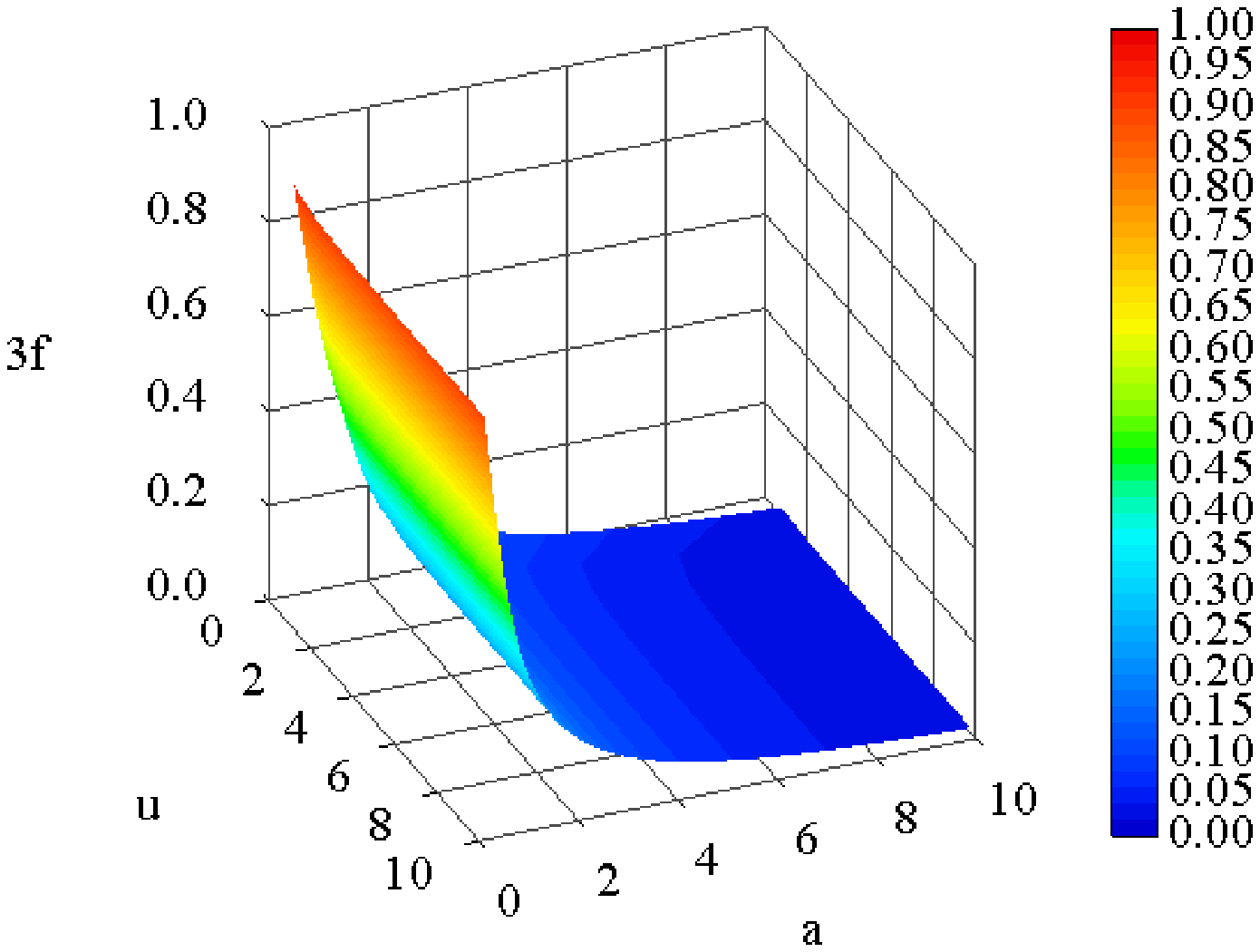}
  \FigureFile(80mm,80mm){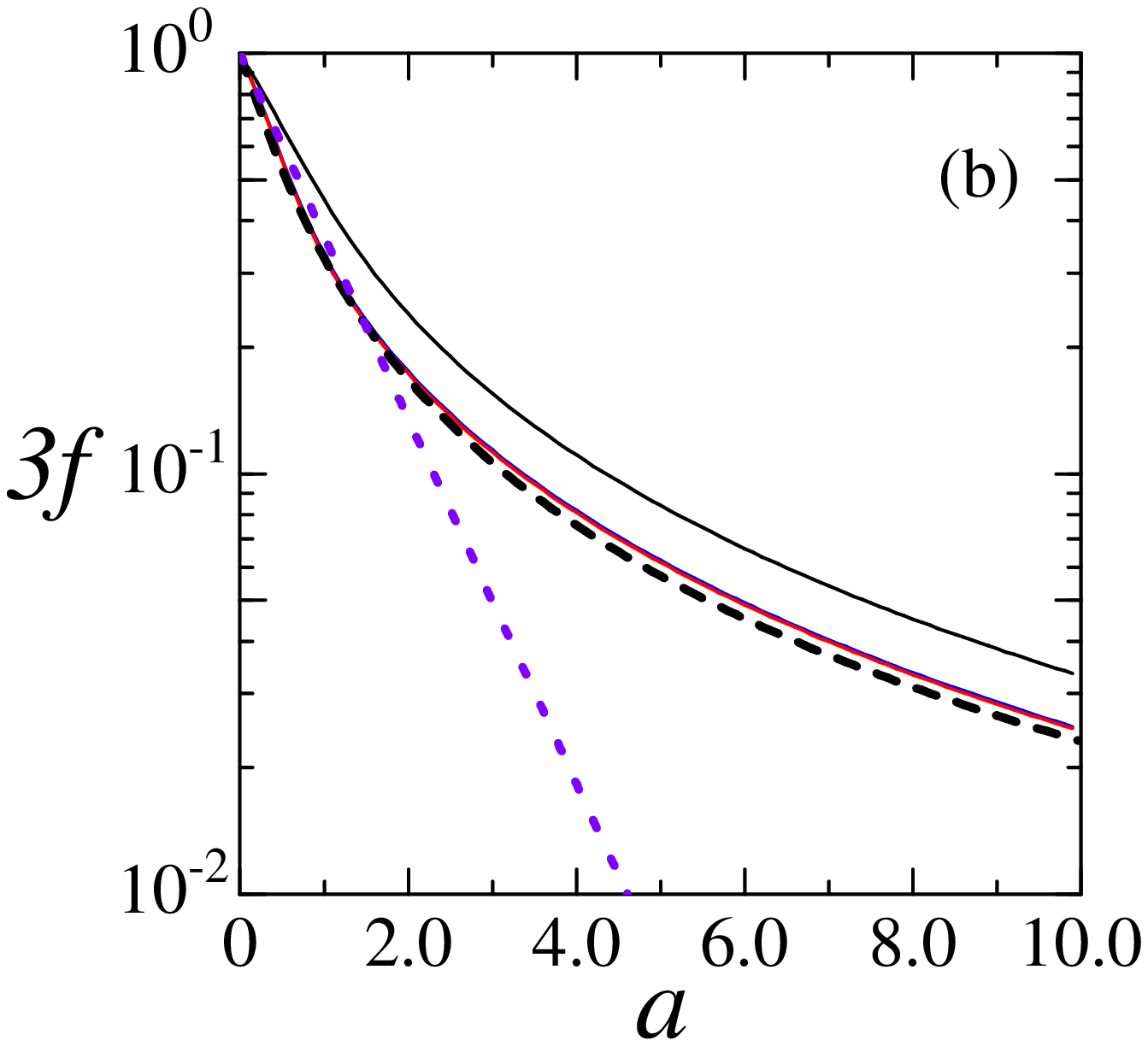}
  \end{center}
\caption{
Behavior of the relativistic variable Eddington factor $f(u, -du/d\tau)$
multiplied by 3 in the case of large $u$.
(a) $3f$ are plotted as a function of the four velocity $u_0$ ($x$ axis)
and the logarithmic four-velocity gradient $-d\ln u/d\tau|_0$ ($a$ axis).
(b) $3f$ are plotted as a function of
the logarithmic four-velocity gradient $-d\ln u/d\tau|_0$
for three fixed $u_0$ ($=1, 5, 10$; solid curves) from top to bottom,
and one fitting curve is also shown (dashed one),
with an exponential curve (dotted curve).
}
\end{figure}

The behavior of RVEF for large $u_0$ is shown in figure 5 (cf. Fukue 2008b).
In this case RVEF decreases,
as the {\it logarithmic} velocity gradient becomes large, 
as shown in Fukue (2008b).
In figure 5b for fixed $u_0$ ($=1, 5, 10$)
the values of $3f$ are plotted as a function of $-d\ln u/d\tau|_0$
by solid curves;
except for the marginally case of $u_0=1$
other curves overlap each other.
One fitting curve is also plotted
by a thick dashed one, and
the numerical results are well fitted by
\begin{equation}
   3f  \sim  \left( 1+\frac{16}{15}a + a^{1.5} \right)^{-1},
\end{equation}
where
\begin{equation}
a=-d\ln u/d\tau|_0.
\end{equation}
Although in Fukue (2008b)
we have used an exponential fitting curve
(the dotted curve in figure 5b),
the present power law type is well fitted at large $a$.

Mixing the small and large $u_0$ cases,
and considering several other patterns,
the relativistic variable Eddington factor in the uniform field
is well fitted by
\begin{equation}
   3f  \sim  \left[ 1+
\frac{16}{15}\left(-\frac{1}{\gamma}\frac{du}{d\tau}\right)
+\left(-\frac{1}{\gamma}\frac{du}{d\tau}\right)^{1.6-2}
               \right]^{-1}.
\end{equation}

We here assumed that the comoving intensity is uniform,
the above result is only due to the redshift effect,
similar to the previous studies (Fukue 2008a, b).
%Indeed, the comoving radiation field
%also decreases as the logarithmic velocity gradient increases
%due to the redshift between the comoving observer
%and the one-tau photo-oval walls.

\begin{figure}
  \begin{center}
  \FigureFile(80mm,80mm){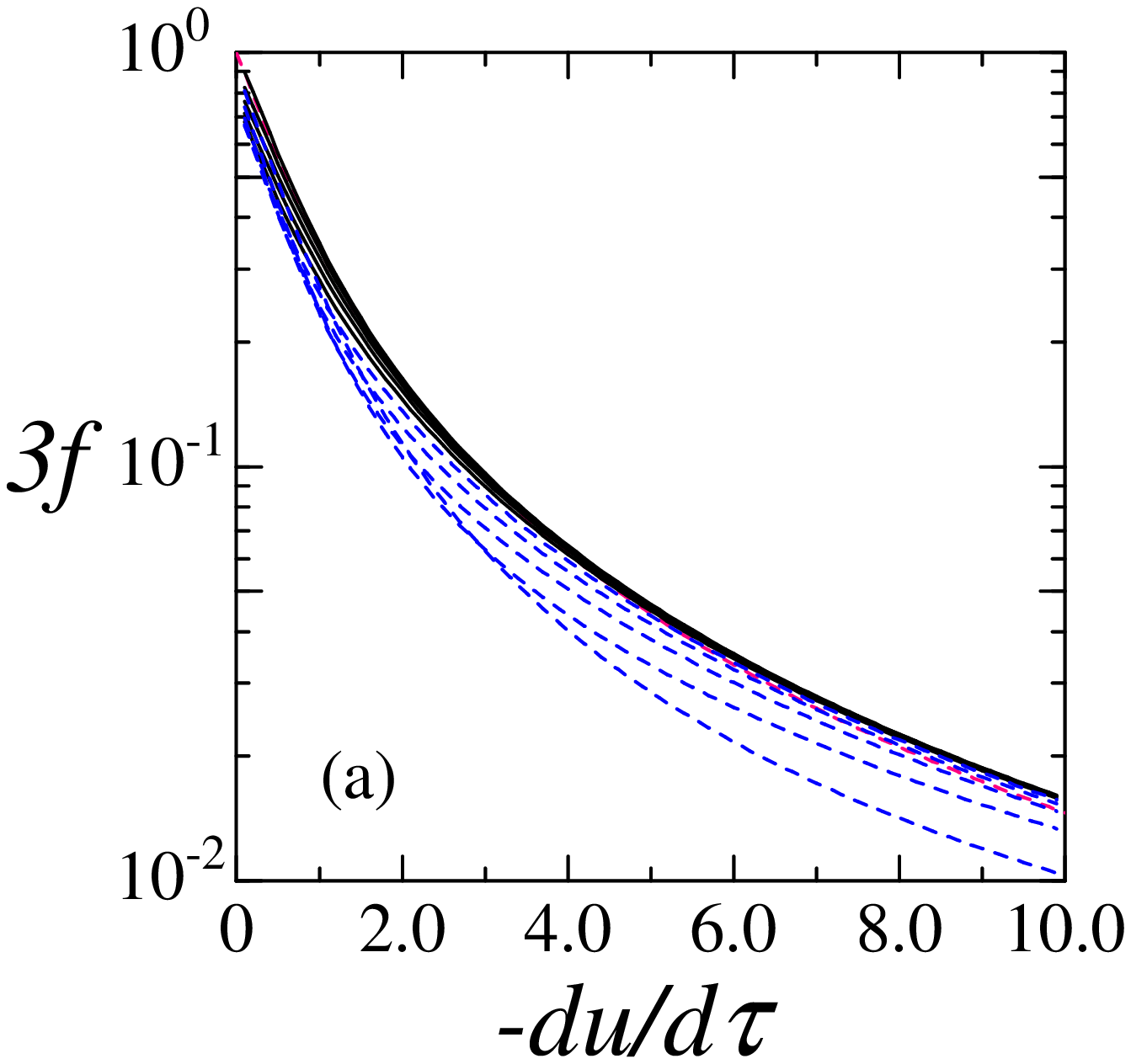}
  \FigureFile(80mm,80mm){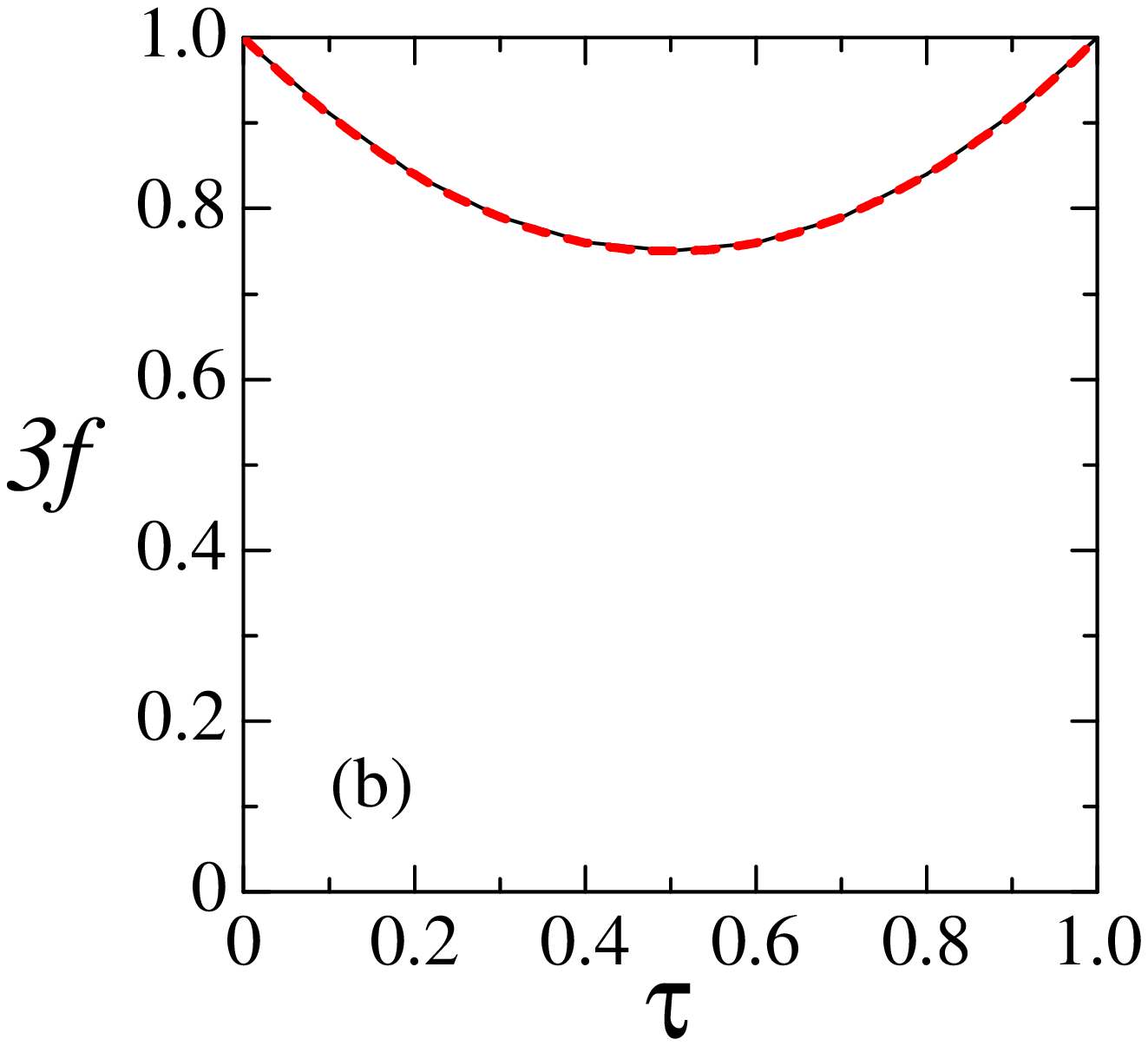}
  \end{center}
\caption{
Behavior of the relativistic variable Eddington factor $f(u, -du/d\tau)$
multiplied by 3 for small $\tau$.
(a) 
The dependence on the velocity gradient is shown
for several values of $\tau_0$
 ($=0.1, 0.2, 0.3, 0.4, 0.5$ from bottom to top of dashed curves)
 ($=0.6, 0.7, 0.8, 0.9, 1.0$ from bottom to top of solid curves)
for $u_0=0.1$.
(b)
The dependence on the optical depth is shown
in the case of $u_0=0.1$ and $-du/d\tau|_0=0$.
One fitting curve is also shown (dashed curve),
}
\end{figure}

Here, we briefly examine the case of small optical depth,
a photo-vessel (figure 1).
That is, as was stated, 
the one-tau region is open towards a downstream direction
to be a photo-vessel for $\tau_0<1$ [equation (\ref{oval})].

The behavior of RVEF for small $\tau$ is shown in figure 6.
In figure 6a 
the values of $3f$ are plotted as a function of $-du/d\tau|_0$
for several values of $\tau_0$
by dashed curves ($\tau_0=0.1, 0.2, 0.3, 0.4, 0.5$
 from bottom to top)
and
by solid curves ($\tau_0=0.6, 0.7, 0.8, 0.9, 1.0$
 from bottom to top)
for $u_0=0.1$.
The tendencies are similar, but
there exist weak deviations.

In figure 6b
the value of $3f$ is plotted as a function of $\tau_0$
for $u_0=0.1$ and $-du/d\tau|_0=0$.
Even if there is no velocity gradient,
the value of RVEF changes with the optical depth.
This is understood as follows.
When the optical depth $\tau_0$ is larger than unity (photo-oval),
$f=1/3$ in the uniform radiation field without the velocity gradient.
When the optical depth becomes smaller than unity (photo-vessel),
the Eddington factor also becomes smaller than $1/3$,
since the radiation field is no longer isotropic;
the radiation from the downstream direction vanishes.
When the optical depth approaches zero, however,
the Eddington factor recovers to $1/3$,
since the Eddington factor above the infinite plane is $1/3$.
The dependence on the optical depth in figure 6b
is well fitted by
\begin{equation}
   f \sim \frac{1}{3} \left( \tau^2 - \tau + 1 \right),
\end{equation}
which is shown by a thick dashed curve in figure 6b,
without any significant error.

\subsection{Weakly Non-Uniform Case}

Next, we examine the optical depth effect,
where the background radiation field is not uniform.
We assume that the comoving radiation field is the Milne-Eddington type:
\begin{equation}
   I (\tau, \mu) = \frac{3}{4} \bar{I} \left( \tau+\frac{2}{3}+\mu \right).
\end{equation}
Here, $\tau$ is the optical depth,
which is related to the observer's quantities by equation (\ref{oval}),
and $\mu$ is the direction cosine in the comoving frame,
which is related to the observer's quantities by
\begin{equation}
   \mu = - \frac{\cos\theta + \Delta\beta}{1+\Delta\beta \cos\theta},
\label{aberration}
\end{equation}
if we consider the aberration effect between the observer
and the radiation site on the photo-oval in the comoving frame.

Using this intensity distribution in the comoving frame,
we easily compute the radiation field and the Eddington factor.
The numerical results are shown in figure 7.
In figures 7a and 7b
the relativistic variable Eddington factors $f(u, -du/d\tau)$
multiplied by 3 are plotted in the parameter space
for the uniform case (figure 7a) and
for the Milne-Eddington case (figure 7b).

\begin{figure}
  \begin{center}
  \FigureFile(80mm,80mm){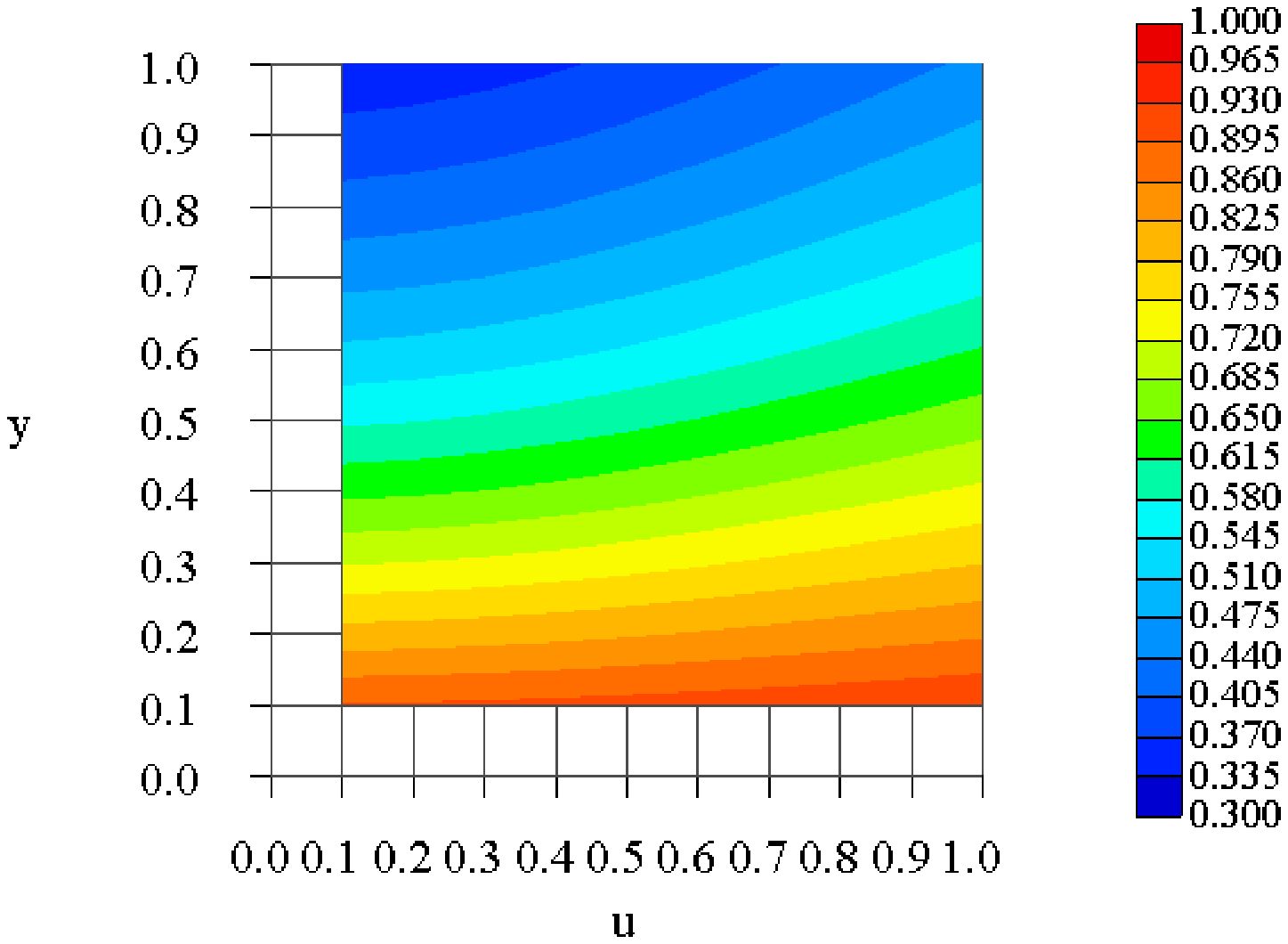}
  \FigureFile(80mm,80mm){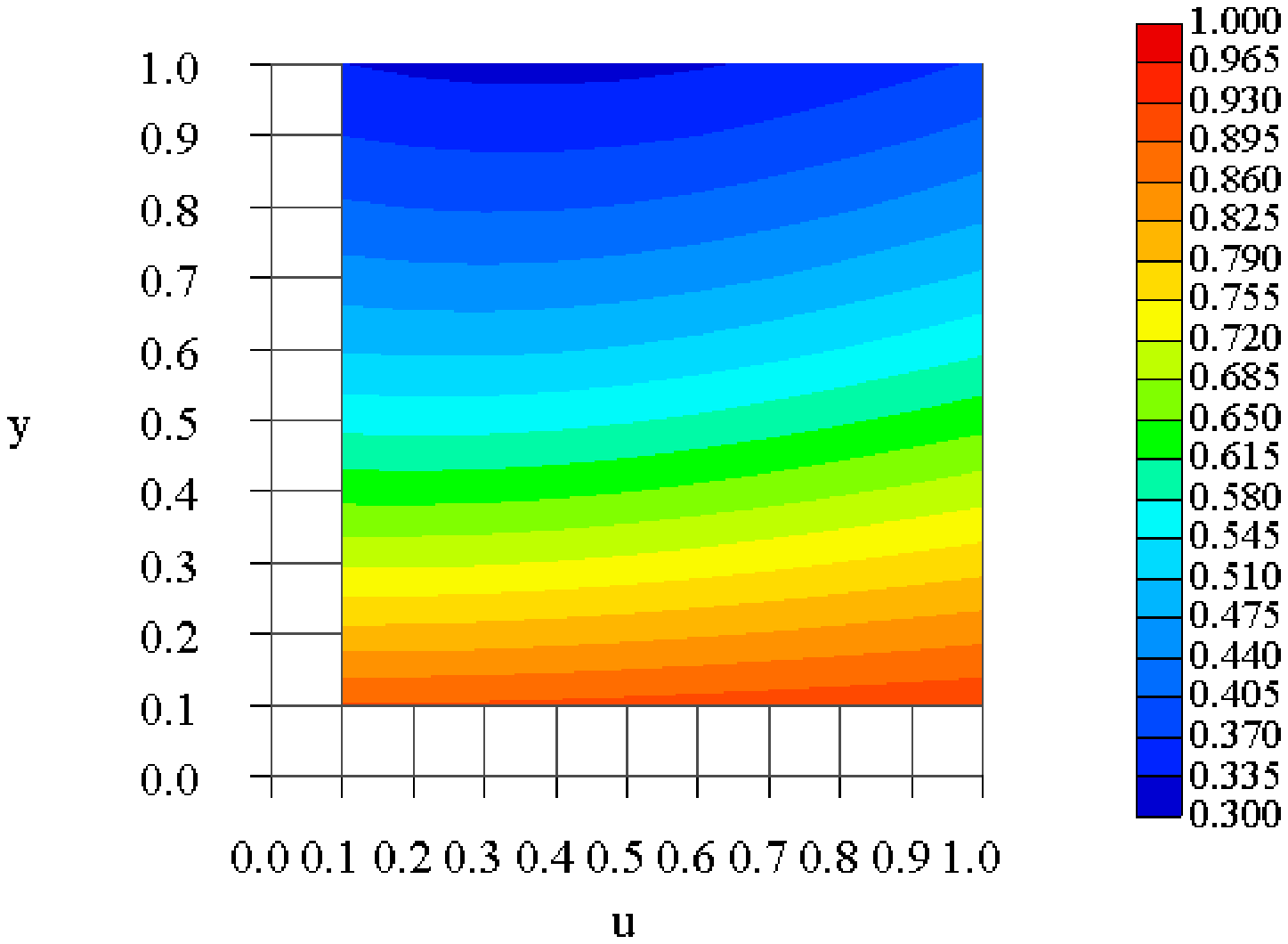}
  \end{center}
\caption{
Behavior of the relativistic variable Eddington factor $f(u, -du/d\tau)$
multiplied by 3.
It is plotted as a function of the four velocity $u_0$ ($x$ axis)
and the four-velocity gradient $-du/d\tau|_0$ ($y$ axis)
in (a) the uniform case and in (b) the Milne-Eddington case
for $\tau_0=1$.
}
\end{figure}

As is seen in figure 7,
the distribution of RVEF of the Milne-Eddington case
is not so much different from that of the uniform case.
For larger $u_0$
the difference becomes much small.
Hence,
the weak non-uniformity or weak anisotropy
does not significantly affect the values of RVEF.

\subsection{Strongly Anisotropic Case}

Thirdly,
we investigate the effect of the relativistic speed itself.
Although there may be various kinds of anisotropy,
we consider the simple limiting case;
we assume that the radiation field in the inertial frame is uniform.
Then, the comoving radiation field is expressed as
\begin{equation}
   I (\tau, \mu, \beta) =
        \frac{1}{\left[ \gamma \left( 1+\beta\mu \right) \right]^4}
                   \bar{I},
\end{equation}
where $\beta$ is the flow speed in units of the speed of light,
$\gamma$ the Lorentz factor, 
and $\mu$ the direction cosine in the comoving frame
given by equation (\ref{aberration}).
Thus, due to the aberration effect
the comoving radiation field becomes strongly anistropic.
Namely,
when the radiative intensity in the comoving frame is uniform,
the intensity in the inertial frame becomes anisotropic
in the sense that the observer in the inertial frame receives
the strong radiation from the upstream direction.
When the radiative intensity in the inertial frame is uniform,
on the other hand,
the intensity in the comoving frame becomes anisotropic
in the sense that the comoving observer receives
the strong radiation from the downstream direction.

The numerical results of this anisotropic case
are shown in figure 8.
In figure 8a
the relativistic variable Eddington factor $f(u, -du/d\tau)$
multiplied by 3 is plotted in the parameter space,
whereas
it is plotted as a function of $u_0$ with fitting curves
in figure 8b.

\begin{figure}
  \begin{center}
  \FigureFile(80mm,80mm){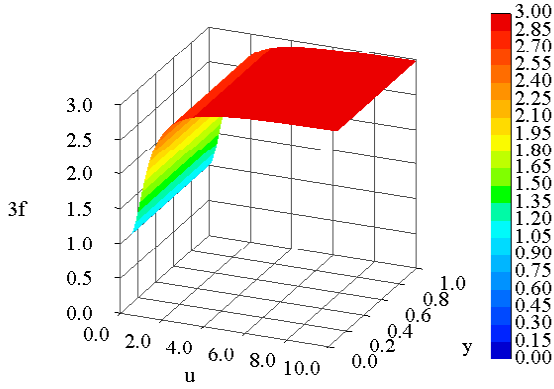}
  \FigureFile(80mm,80mm){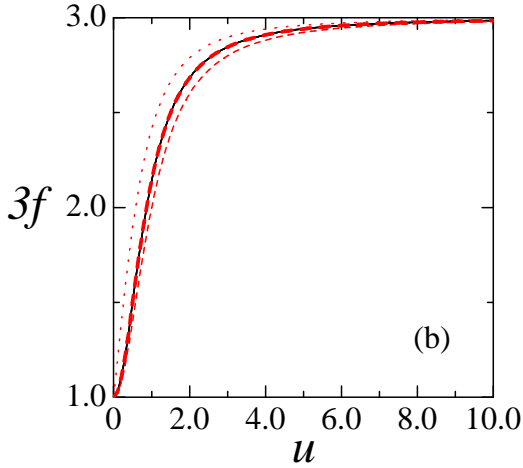}
  \end{center}
\caption{
Behavior of the relativistic variable Eddington factor $f(u, -du/d\tau)$
multiplied by 3 in the strongly anisotropic case.
(a) $3f$ are plotted as a function of the four velocity $u_0$ ($x$ axis)
and the four-velocity gradient $-du/d\tau|_0$ ($y$ axis).
(b) $3f$ are plotted as a function of the four velocity $u_0$
for three fixed $du/d\tau|_0$ ($=0, 0.5, 1.0$; solid curves),
and one fitting curve is also shown (dashed curve),
with two comparisons (thin dashed and dotted curves).
}
\end{figure}

The behavior of RVEF shown in figure 8a is quite impressive,
since it seems that $f$ does not depend on the velocity gradient,
but depends only on the velocity.
Indeed, in figure 8b
the values of $3f$ are plotted as a function of $u_0$
for several fixed $-du/d\tau|_0$ ($=0, 0.5, 1$) by solid curves,
and all curves overlap each other.
One fitting curve is also plotted
by a thick dashed one,
and the numerical results are well fitted by
\begin{equation}
   3f = \frac{3(1+4u^2)}{3+4u^2}=\frac{3(1+3\beta^2)}{3+\beta^2}.
\label{best}
\end{equation}
As comparisons,
RVEF proposed in Fukue (2006) are also plotted
by the thin dashed one ($3f=1+2\beta^2$) and
the thin dotted one ($3f=1+2\beta$).

As is seen in figure 8b,
equation (\ref{best}) reproduces the numerical results very well.
This is interpreted as follows.
When the radiation intensity in the inertial frame is uniform,
the radiation energy density $E_{\rm lab}$,
the radiative flux $F_{\rm lab}$, and
the radiation pressure $P_{\rm lab}$ in the inertial
(laboratory) frame
become, respectively,
\begin{eqnarray}
   cE_{\rm lab} &=& 4\pi \bar{I},
\\
    F_{\rm lab} &=& 0,
\\
   cP_{\rm lab} &=& \frac{4}{3}\pi \bar{I},
\end{eqnarray}
for $\tau_0>1$.
Hence, from the Lorentz transformation
the radiation energy density $E_{\rm co}$,
the radiative flux $F_{\rm co}$, and
the radiation pressure $P_{\rm co}$ in the comoving frame
become, respectively,
\begin{eqnarray}
   cE_{\rm co} 
   &=& \gamma^2 \left(cE_{\rm lab}-2\beta F_{\rm lab}+\beta^2 cP_{\rm lab} \right)
\nonumber \\
   &=& 4\pi \bar{I} \gamma^2 \left( 1+\frac{1}{3}\beta^2 \right),
\\
    F_{\rm co}
   &=& \gamma^2 \left[(1+\beta^2)F_{\rm lab}-\beta(cE_{\rm lab}+cP_{\rm lab}) \right]
\nonumber \\
   &=& -4\pi \bar{I} \frac{4}{3}\gamma^2 \beta,
\\
   cP_{\rm co}
   &=& \gamma^2 \left(\beta^2cE_{\rm lab}-2\beta F_{\rm lab}+cP_{\rm lab} \right)
\nonumber \\
   &=& 4\pi \bar{I} \gamma^2 \left( \beta^2+\frac{1}{3} \right).
\end{eqnarray}
Thus, the relativistic variable Eddington factor analytically becomes
\begin{equation}
   f \equiv \frac{P_{\rm co}}{E_{\rm co}} = \frac{1+3\beta^2}{3+\beta^2}.
\label{ana}
\end{equation}

%%%%%%%%%%%%%  CONCLUDING REMARKS  %%%%%%%%%%%%%%%%%%%%%%%%%%%%

\section{Concluding Remarks}

In this paper, 
we have derived the relativistic variable Eddington factor 
for a plane-parallel flow that is accelerating in the vertical direction;
we have introduced the one-tau photo-oval observed by the comoving observer,
and then calculated the comoving radiation field and the Eddington factor
for several situations,
and found that
the Eddington factor in the relativistic regime
generally depend on the optical depth,
the flow velocity, and the velocity gradient.

Of these,
the dependence of RVEF on the optical depth
is weak
similar to the non-relativistic plane-parallel atmosphere.
As for the dependence on the velocity gradient, on the other hand,
RVEF generally decreases as the velocity gradient increases.
When the velocity gradient is small,
RVEF linearly decreases with the velocity gradient,
as proved in Fukue (2008a).
When the velocity gradient becomes large,
RVEF decreases in the power law manner
of the logarithmic velocity gradient (cf. Fukue 2008b).
Finally, as for the dependence on the relativistic speed,
RVEF can increases to be on the order of unity,
if the comoving intensity is sufficiently anisotropic.

We have considered the case
of the plane-parallel vertical flow.
In the spherical flow, on the other hand,
there exists a geometrical dilution effect,
and the situation can be somewhat different,
as $r^2$ terms appear (e.g., Peraiah 2002).
The present approach, however, can be 
extended to treat also the spherical case
and we will examine it in future work.

The treatment of the relativistic Eddington factor
presented in this paper may turn out to be applicable
in various aspects of relativistic astrophysics with radiation transfer;
i.e., 
black-hole accretion flows with supercritical accretion rates,
relativistic jets and winds driven by luminous central objects,
relativistic explosions including gamma-ray bursts,
neutrino transfers in supernova explosions,
and various events occured in the proto universe
(cf. Fukue 2008b).
The treatment of the relativistic Eddington factor
presented in this paper will help in investigating
these types of astrophysical problems
that involve the solution of the fully relativistic
radiation hydrodynamical equations.

\vspace*{1pc}

The author would like to express his sincere thanks to M. Umemura,
T. Koizumi, and C. Akizuki
for their enlightening and stimulated discussions.
%Valuable comments to improve the original manuscript
% from anonymous referees are also gratefully acknowledged.
This work has been supported in part
by a Grant-in-Aid for Scientific Research (18540240 J.F.) 
of the Ministry of Education, Culture, Sports, Science and Technology.

%\noindent

\end{document}